\documentclass{article}
\usepackage[a4paper, total={6.5in, 10in}]{geometry}
\usepackage{cite}
\usepackage{amsmath,amssymb,amsfonts}
\usepackage{algorithmic}
\usepackage{graphicx}
\usepackage{subcaption}
\usepackage{textcomp}
\usepackage[table]{xcolor}
\usepackage{soul}
\usepackage{todonotes}
\usepackage{listings}
\usepackage{booktabs}
\usepackage{multirow}
\usepackage{hyperref}

\def\BibTeX{{\rm B\kern-.05em{\sc i\kern-.025em b}\kern-.08em
    T\kern-.1667em\lower.7ex\hbox{E}\kern-.125emX}}

\definecolor{codegreen}{rgb}{0,0.6,0}
\definecolor{codegray}{rgb}{0.5,0.5,0.5}
\definecolor{codepurple}{rgb}{0.58,0,0.82}
\definecolor{backcolour}{rgb}{0.95,0.95,0.92}

\lstdefinestyle{cppstyle}{
    backgroundcolor=\color{backcolour},   
    commentstyle=\color{codegreen},
    keywordstyle=\color{magenta},
    numberstyle=\tiny\color{codegray},
    stringstyle=\color{codepurple},
    basicstyle=\ttfamily\footnotesize,
    breakatwhitespace=false,         
    breaklines=true,                 
    captionpos=b,                    
    keepspaces=true,                 
    numbers=left,                    
    numbersep=3pt,                  
    showspaces=false,                
    showstringspaces=false,
    showtabs=false,                  
    tabsize=2
}

\newcommand{\ra}[1]{\renewcommand{\arraystretch}{#1}}

\begin{document}

\title{Porting HPC Applications to AMD Instinct$^\text{TM}$ MI300A Using Unified Memory and OpenMP\textregistered}

\author{$^\mathsection$Suyash Tandon, $^\mathsection$Leopold Grinberg, $^\mathsection$Gheorghe-Teodor Bercea, $^\mathsection$Carlo Bertolli, \\$^\dagger$Mark Olesen, $^\ddagger$Simone Bn\`a, and $^\mathsection$Nicholas Malaya \\ 
        $^\mathsection$\textit{Advanced Micro Devices, Inc.}, Austin, U.S.A. \\
        $^\dagger$ \textit{OpenCFD Ltd.}, Berkshire, U.K.\\
        $^\ddagger$ \textit{SuperComputing Applications and Innovations Department, CINECA}, Bologna, Italy.}
\date{ }
\maketitle

\begin{abstract}
AMD Instinct$^\text{TM}$ MI300A is the world's first data center accelerated processing unit (APU) with memory shared between the AMD ``Zen 4" EPYC$^\text{TM}$ cores and third generation CDNA$^\text{TM}$ compute units. A single memory space offers several advantages: i) it eliminates the need for data replication and costly data transfers, ii) it substantially simplifies application development and allows an {\it incremental} acceleration of applications, iii) is easy to maintain, and iv) its potential can be well realized via the abstractions in the OpenMP\textregistered\ 5.2 standard, where the host and the device data environments can be unified in a more performant way. In this article, we provide a blueprint of the APU programming model leveraging unified memory and highlight key distinctions compared to the conventional approach with discrete GPUs. OpenFOAM\textregistered, an open-source C++ library for computational fluid dynamics, is presented as a case study to emphasize the flexibility and ease of offloading a full-scale production-ready application on MI300 APUs using directive-based OpenMP programming.

\end{abstract}

% ==============================================
% SECTION I: Introduction
% ==============================================
\section{Introduction}
High-performance computing (HPC) node architecture designs are driven primarily by power management considerations and are increasingly reliant on high degrees of fine-grained parallelism~\cite{b1,b2}. Top500 list~\cite{b3} mentions over 150+ known systems that use accelerator/co-processor technology. Although these heterogeneous architectures provide a high performance per-watt advantage in comparison to the traditional homogeneous CPU-based systems~\cite{b4}, porting, tuning, and maintaining scientific applications with millions of code lines can be tedious and challenging. Additionally, the desire for user applications to run efficiently on a variety of accelerators often renders non-portable programming impractical as application developers may lack detailed knowledge of a specific accelerator and its hardware intricacies as well as growing concerns around code maintainability and duplication~\cite{b5}. Directive-based programming models allow developers to insert compiler directives into a code region to automatically generate parallel code for the target system. Two popular directive-based programming models that have been widely adopted are OpenMP\textregistered~\cite{b6} and OpenACC~\cite{b7}.

OpenMP\textregistered\ $4.0$ and beyond have made a paradigm change to support heterogeneous systems and leverage accelerators like GPUs with the ability to offload computations to accelerators~\cite{b8}. In more recent releases~\cite{b9}, new features to manage memory on heterogenous systems have been added with full support for accelerator devices. Increasing compiler support and optimizations have enabled numerous case studies and user experiences of OpenMP\ target offloading of in-house applications~\cite{b10}, mini-apps~\cite{b11}, and benchmarks~\cite{b12}. However, the simplicity of the example codes presented in these case studies often creates a challenge when translating and implementing OpenMP target offloading in production-ready applications. 

Managing memory and data in a performant way is one of the primary challenges in porting and optimizing complex applications for systems with discrete memory spaces and compute devices. Some recent work~\cite{b12} analyzed using unified memory, enabling the CPU and GPU to access the same memory. Before the introduction of unified memory, {\it host} and {\it devices} would use separate memory spaces and as a result, communication between CPU and GPUs had to be managed by programmers explicitly. 

From the coding perspective, unified virtual memory addressing supported on systems with discrete physical memories of compute devices provides a unified view of the data and reduces the programmability complexity. However, low performance due to insufficient hardware or system software support for unified virtual memory addressing may render such an approach not practical. For example, excessive page migrations triggered by accessing the same virtual memory by the host and the device can overshadow any performance advantages in using accelerators. Initial work on assessing unified memory on systems with discrete GPUs reported low portability between systems~\cite{b12, b13} and lack of compiler optimizations~\cite{b13} resulting in major bottlenecks. APU offers an unified physical memory, which eliminates the need for page migrations and further advances the performance in fine-grained data sharing by the host and device threads. 

In this paper, we focus on programming on AMD Instinct$^\text{TM}$ MI300A using the latest OpenMP standard, and specifically its support for unified memory. To demonstrate the impact of the alignment of innovation in the hardware and system software we go beyond using simple code examples. Specifically, we present our experience in porting within the framework of OpenFOAM\textregistered, an established open-source C++ library and application suite for computational fluid dynamics that comprises on the order of 1 million lines of code~\cite{b14}. We emphasize the flexibility and ease of offloading a {\it production-ready} application on MI300 APUs using directive-based programming. The contributions of this paper are the following: 
\begin{itemize}
  \item We provide a blueprint of the APU programming model and demonstrate the ease and flexibility of porting codes on MI300A with OpenMP. 
  \item We elaborate our method for incremental acceleration of a production and widely used in industry code---OpenFOAM.
\end{itemize}

The rest of this paper is organized as follows: \S\ref{sec:related-work} highlights some notable work from the past. \S\ref{sec:MI300A} provides background on OpenMP target offloading, high-level details of AMD Instinct$^\text{TM}$ MI300A node architecture and describes some of the key advantages of programming with OpenMP instead of HIP\cite{b15}/CUDA\textregistered\cite{b16}. \S\ref{sec:case-study} presents a case-study to demonstrate the ease and flexibility in porting a production-ready HPC application to MI300A and \S\ref{sec:performance} compares the performance on discrete GPUs and MI300A. Finally, \S\ref{sec:conclusion} provides the main concluding remarks of this paper.

% ==============================================
% SECTION II: Related Work 
% ==============================================
\section{Related Work}
\label{sec:related-work}

Programming models based on compiler directives for offloading to GPUs are gradually becoming a real alternative to programming models based on HIP\cite{b15}/CUDA\cite{b16}. Several production codes such as QMCPACK\cite{b17},  VASP\cite{b18}, ICON\cite{b19}, GenASiS\cite{b20}, etc. employ either OpenACC or OpenMP based acceleration. In addition, OpenMP target offloading to accelerate applications from various scientific domains on discrete GPUs has been the context of several case studies (e.g., Nekbone~\cite{b21}, Lulesh~\cite{b22}, and U.K mini-apps~\cite{b23} among others). The analysis presented in the previous publications can be categorized into compiler optimization, runtime overheads, and data management challenges. Compiler optimizes compute kernels, achieving high performance with OpenMP target offload on CPU and GPU targets when using {\tt teams distribute parallel for} constructs and avoiding the use of explicit {\tt schedules}~\cite{b24,b25}. Other compiler optimization research has been focussed on accelerating user code that exists between the {\tt target} and {\tt parallel} constructs~\cite{b24,b26,b27}. Detailed analysis of OpenMP 4.5 supported by different compilers show runtime overheads during the testing of different features~\cite{b28}. More recently~\cite{b11}, three compilers supporting OpenMP directives for offloading tested discrete GPU compute capabilities, and runtime overheads in LLVM/Clang were identified with suggestions for manual implementation of {\tt acc\_attach} to create data structure on device and find association between host and device addresses. Similarly to using HIP/CUDA programming models, in directive-based programming the data management challenges have been one of the major hurdles in extending the applicability of OpenMP GPU offloading from benchmarks to full-scale production codes. The challenge of dealing with nested data is described in~\cite{b13}, and some code transformations to work with nested structures have been proposed. However, alternative methods to map struct containing pointers~\cite{b29} using managed memory allocations via {\tt cudaMallocManaged} simplify the use of C$++$ objects, and enable use of classes like {\tt std::vector} in OpenMP target  regions. 

Many successful attempts with OpenMP target offloading using unified memory have been presented, however, the majority of work has leveraged benchmarks and mini-apps. Unified memory addressing is not available on all CPU $+$ GPU platforms and its use in some cases has been associated with higher than expected overheads~\cite{b13}. While explicit data management is possible through abstractions using OpenMP runtime API calls, e.g., Kokkos~\cite{b30}, the lack of compiler support can be problematic. In this paper, we focus on directive-based programming using OpenMP, and specifically, on the advantages of the unified memory approach in programming supported by the unified physical memory in the MI300A. 

% ==============================================
% SECTION III: Programming MI300A 
% ==============================================
\section{Programming MI300A with OpenMP}
\label{sec:MI300A}

\begin{figure}[htb]
  \centering
  \begin{subfigure}{0.22\textwidth}
    \includegraphics[width=\textwidth]{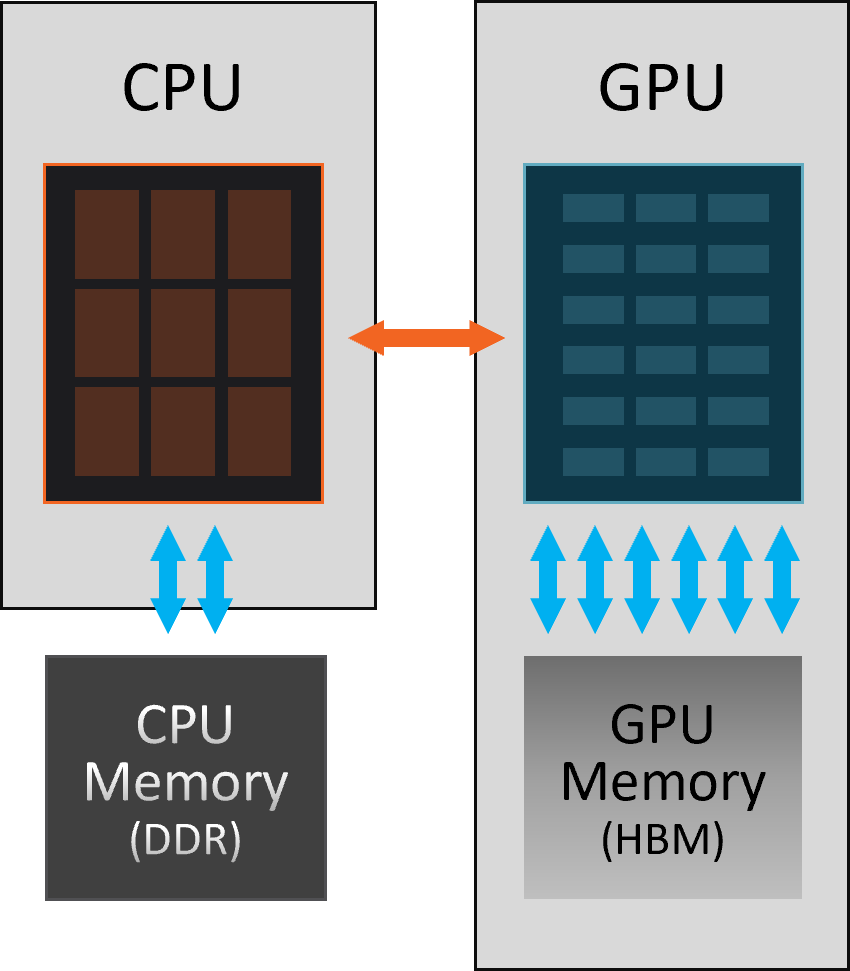}
    \caption{Discrete CPU and GPU.}
    \label{subfig:dgpu}
  \end{subfigure}
  \hspace{0.02\textwidth}
  \begin{subfigure}{0.22\textwidth}
    \includegraphics[width=\textwidth]{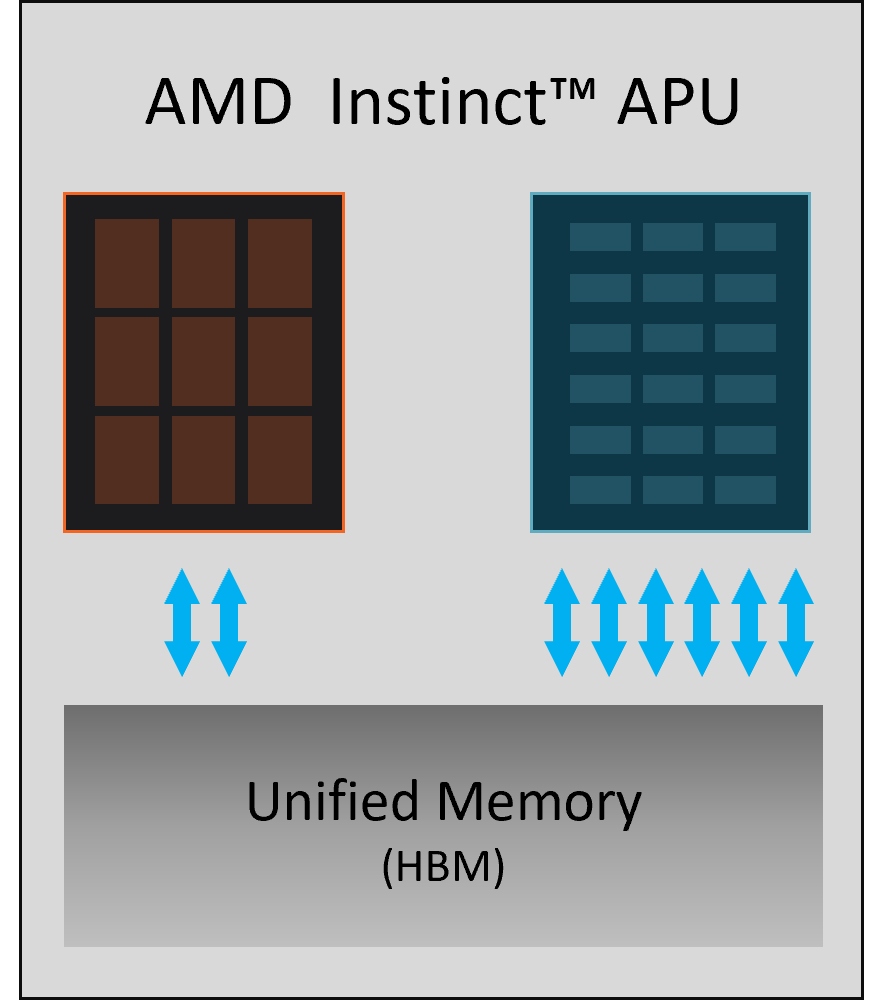}
    \caption{APU.}
    \label{subfig:apu}
  \end{subfigure}
  \caption{Schematic representation of a socket with a discrete CPU and GPU and a socket with APU.}
  \label{fig:dgpu-vs-apu}
\end{figure}

Figure~\ref{fig:dgpu-vs-apu} shows the schematics of a socket with a) a discrete CPU and GPU, and b) an APU. In the APU design, the AMD ``Zen 4" EPYC$^\text{TM}$ cores and third generation CDNA$^\text{TM}$ compute units share the same high-bandwidth memory, which eliminates the need for data replication and does not require a programming distinction between the {\it host} and {\it device} memory spaces. 

High performance, fine-grained sharing of the same virtual and physical memory by the threads running on the CPU and GPU processing elements enables a unified programming framework that provides software developers with the ability to: i) alternate the work on the same data sets between the CPU and the GPU without taking the penalty of data transfers, and ii) write significantly less code. OpenMP specifications include an informational pragma that allows programmers to select {\tt unified\_shared\_memory} as a requirement for the execution of a program. ROCm$^\text{TM}$ supports the requirement and its implementation is based on AMDGPU Unified Memory support.

\begin{lstlisting}[caption={Simple example of directive-based code acceleration with OpenMP and {\tt unified\_shared\_memory}.}, language=C++, label={lst:usm-data}]
#define N (1024*100)

#pragma omp requires unified_shared_memory

int main() {
  double *a = new (std::align_val_t(__STDCPP_DEFAULT_NEW_ALIGNMENT__)) double[N];
  double *b = new (std::align_val_t(__STDCPP_DEFAULT_NEW_ALIGNMENT__)) double[N];
  double k = 0.0;
  
  // fill a and b arrays and constant k from file
  fileIO(a, b, &k, N);

  #pragma omp target teams distribute parallel for
  for(int i = 0; i < N; i++)
    b[i] += a[i] * k;
}
\end{lstlisting}

Consider the program in listing~\ref{lst:usm-data}. A single OpenMP host thread executes code in the main function and it launches a GPU kernel execution when using the pragma {\tt target}. First, it allocates heap memory using the new operator. Note that we use C++17 semantics to control memory alignment via the code compilation. Also note that while we use {\tt new} to allocate memory, on MI300A any memory allocator including {\tt hipMalloc} will allocate unified memory, i.e., memory accessible by any compute element.  Second, it populates the arrays with input data in a separate function. Then it launches a GPU kernel  using the arrays and a constant value. The program does not contain any OpenMP data mapping because it specified the requirement {\tt unified\_shared\_memory}, which ``makes map clauses optional on target constructs''\cite{b9}. Following OpenMP specification, the compiler maps pointers {\tt a} and {\tt b} as zero-sized array sections, with the pointers as the base. That is, the compiler adds an implicit map clause as following {\tt map(a[:0], b[:0])}. The effect of such mapping is that the contents of {\tt a} and {\tt b} are passed by value as GPU kernel arguments. Inside the GPU kernel, threads access {\tt a} and {\tt b} pointers from the kernel function parameters. When accessing addresses based on {\tt a} and {\tt b} pointers (e.g., {\tt a[0]}, {\tt a[1]}, etc.), GPU threads emit loads and stores with the host pointer values as the base address.

The ROCm software stack and its AMDGPU hardware support provide an implementation of such host-pointer addressing from GPU kernels using its Unified Memory support. OpenMP programmers are oblivious of how the GPU support functions, and they only need to know that {\tt a} and {\tt b} are passed by-value to the GPU kernel.

In addition, using functions inside a GPU kernel requires a different programming effort because unlike data that can be accessed from any device on AMDGPU-based systems, code that has been compiled for the host cannot be executed on a GPU device. OpenMP programmers need to mark all functions that might be called inside a GPU kernel using the pragma {\tt declare target}. The compiler will generate a device (and host) code for all those functions marked with the pragma and replace calls to those functions inside GPU kernels to the corresponding GPU implementations. Marking functions for GPU code generation is easily achieved in simple applications that do not make use of third-party libraries inside target regions.

In general applications, developers are interested to employ third-party library calls in target regions, however, they do not normally have access to their implementation, nor should they be expected to modify them.
Consider the program in listing~\ref{lst:implicit}, which is a simplified coding of the daxpy algorithm using C++ STL vectors.
\begin{lstlisting}[caption={Example of directive-based offloading with nested data and implicit code generation for {\tt std::vector} class methods.}, language=C++, label={lst:implicit}]
#include <vector>
#pragma omp requires unified_shared_memory

template<typename T>
void daxpy(T da, vector<T> dx, vector<T> dy) {
  #pragma omp target teams distribute parallel for
  for(auto i = 0; i < dx.size(); i++)
    dy[i] = dy[i] + da*dx[i];
}

int main() {
  std::vector<double> dx, dy;
  double da;
  // input da, dx, and dy from file
  fileIn(da, dx, dy);
  
  daxpy(da, dx, dy);
}

\end{lstlisting}
The program uses {\tt unified\_shared\_memory} to enable GPU threads to access variables declared and allocated on host memory. This includes program stack variables {\tt dx} and {\tt dy} and the host pointers they encapsulate for the vector data, which is allocated on the heap. The target region uses the {\tt size} function of C++ vectors, as well as the {\tt []} operator to access elements in the vector data buffers. The OpenMP compiler identifies the function and operator as needed to correctly generate code for the GPU, and  if their implementations are visible in the {\tt vector} header file, it generates their code in the GPU binary.

This support is not available when the implementation of a function, called from within a target region, is not visible when compiling the file that contains the target region. This often happens when the implementation is stored in a separate C++ implementation source file: the compiler can only see the declaration of the function in the included header file, but not its definition in the backing implementation file. In this case, the user is required to manually modify the source code of the implementation file to mark the function with the {\tt declare target} pragma. In the example above, this is not necessary as the implementation of {\tt size} and {\tt []} is provided in the {\tt vector} header file.

It is worth mentioning that the ROCm software stack provides an evolved API compared to what was described in a previous paper\cite{b31}, where instead of special allocators the default C++ vector allocator can now be used. This means that memory allocated using standard OS allocators such as mmap, sbrk, and similar, can be accessed on AMDGPUs and supported by the ROCm unified memory facility.

% ==============================================
% SECTION IV: Case study - OpenFOAM
% ==============================================
\section{Case Study: OpenFOAM}
\label{sec:case-study}
OpenFOAM is an established and wide-spread open-source C++ library~\cite{b14} that uses object-oriented programming and tensor algebra to implement mathematical models for computational continuum mechanics with special focus for computational fluid dynamics (CFD). In OpenFOAM, the computational domain is divided into a set of discrete volumes $\displaystyle\partial V_i$ that fill the computational domain $D$ without overlap, i.e., $\displaystyle\cup_i\partial V_i=D$ and $\displaystyle \cap_i\partial V_i=\varnothing$. The fluid-flow equations are then volume-integrated over each individual finite volume $\partial V_i$. Parallelization on distributed systems uses MPI for inter-process communication, where the domain is split into $N$ subdomains and neighboring processors are joined with processor-to-processor boundary conditions that are responsible for exchanging information during the simulation.

From the code organization and architecture perspective, OpenFOAM makes extensive use of templated classes, macros, and data encapsulation. Such choice hides many lower-level details of the code, for example,  implementation of the basic tensorial classes, operator overloading, enabling parallel communication, from view at the higher levels and makes it easy for user to implement new continuum mechanics models~\cite{b14}. At the same time, such a code architecture makes porting $O$(1M) lines of the CPU-code to GPUs quite challenging. Efforts to port OpenFOAM to discrete GPUs have been made in the past, and resulted in either stand-alone forks~\cite{b32} or enabling GPU-acceleration by interfacing OpenFOAM with other third-party libraries like PETSc~\cite{b33}, Ginkgo~\cite{b34}, etc. All of those efforts are based on data duplication and managing two discrete memory spaces---one for the GPU and another for the CPU. The third-party packages can themselves be quite large---both PETSc and Ginkgo contain $O$(1M) code lines, with further complex dependencies on libraries such as {\it hipSparse} or {\it cuSparse}.

As an example, the code snippet in listing~\ref{lst:openfoam-simple} is the main part of the code, {\tt simpleFoam}, which is a steady-state solver for incompressible, turbulent flows solving the Navier-Stokes equations with turbulence modeling. In the predictor-corrector SIMPLE algorithm~\cite{b35} (listing~\ref{lst:openfoam-simple}), the pressure and velocity fields are decoupled and solved iteratively as follows:
\begin{enumerate}
  \item Pressure solution from a previous time-step is used as an initial guess and the momentum equation is solved in line 10 to a predefined tolerance to give an approximate velocity field,
  \item The pressure Poisson equation is formulated on line 19 with the divergence of the partial velocity flux as a source term and solved to give a new estimate of the pressure field on line 23. A new set of conservative fluxes is obtained from the pressure equation,
  \item The corrected pressure field is used in an explicit correction to the velocity field on line 32,
  \item Solve the transport equations; turbulent quantities on line 37.
\end{enumerate}

\begin{lstlisting}[caption={Code snippet from {\tt simpleFoam.C} showing different stages of computations as per SIMPLE algorithm.}, language=C++, label={lst:openfoam-simple}]
  while (simple.loop())
  {
    Info<< "Time = " << runTime.timeName() << nl << endl;
    // --- Pressure-velocity SIMPLE corrector
    {
      // Momentum predictor
      ...
      if (simple.momentumPredictor())
      {
          solve(UEqn == -fvc::grad(p));
          fvOptions.correct(U);
      }
      ...
      // Non-orthogonal pressure corrector loop
      while (simple.correctNonOrthogonal())
      {
          fvScalarMatrix pEqn
          (
              fvm::laplacian(rAtU(), p) == fvc::div(phiHbyA)
          );

          pEqn.setReference(pRefCell, pRefValue);
          pEqn.solve();

          if (simple.finalNonOrthogonalIter())
          {
              phi = phiHbyA - pEqn.flux();
          }
      }
      ...
      // Momentum corrector
      U = HbyA - rAtU()*fvc::grad(p);
      U.correctBoundaryConditions();
      fvOptions.correct(U);
    }

    laminarTransport.correct();
    turbulence->correct();
    ...
    }
\end{lstlisting}

\begin{figure*}
  \centerline{\includegraphics[width=\textwidth, trim=0 2cm 0 0, clip]{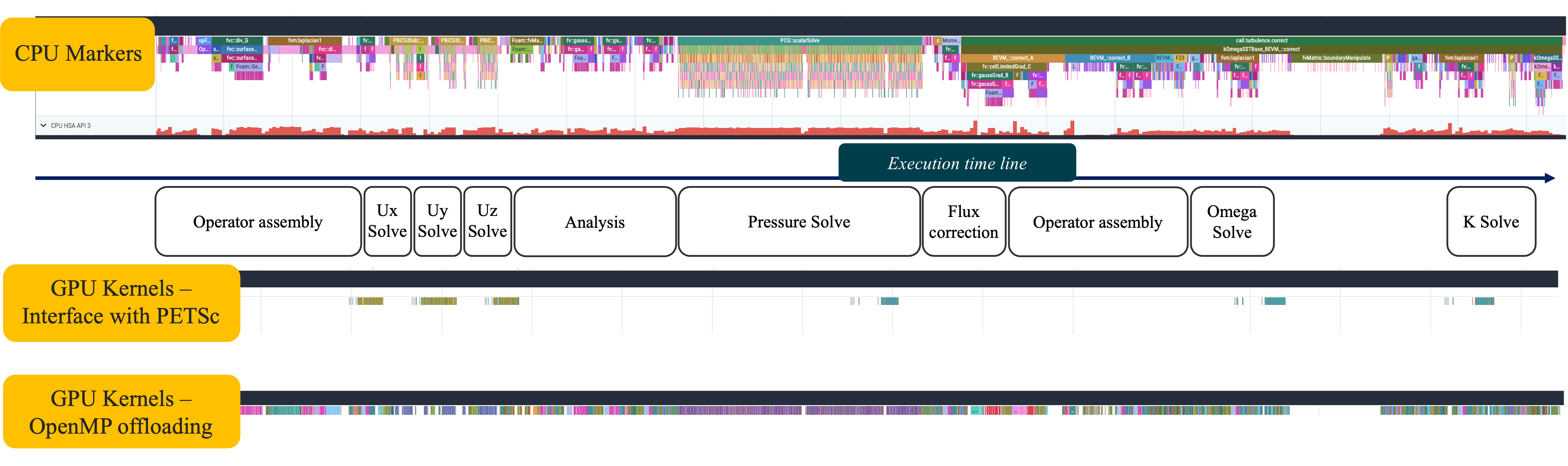}}
  \caption{Markers for the CPU-execution (top half) and trace of the GPU kernels offloaded (bottom-half) with PETSc interface show the execution of the {\tt HPC\_motorbike} benchmark (Large with 34 M cells) with {\tt simpleFoam} solver for a single time step $t_i$. Identifiers for the different stages of computations are marked in the {\it Execution time line}.}
  \label{fig:openfoam-time-step-details}
\end{figure*}

The {\tt while} loop runs for a desired time $T=\sum t_i$, and within an individual time step $t_i$ the inner pressure-correction (steps $2$--$3$) can be iterated as many times as necessary for convergence. The {\tt HPC\_motorbike} benchmark (Large with 34 M cells)~\cite{b36} uses {\tt simpleFoam} solver and the top-half of Figure~\ref{fig:openfoam-time-step-details} uses markers to provide a more detailed view of the computations in an individual time-step. For simplicity, {\it Execution time line}, is provided which helps connect the markers with the code in listing~\ref{lst:openfoam-simple}. Bottom half of Figure~\ref{fig:openfoam-time-step-details} shows a trace of the {\tt HPC\_motorbike} benchmark configured with PETSc. In comparison to the CPU-executions, PETSc interface offloads only the {\tt KSPSolve}~\cite{b33} kernels to GPUs and a significant fraction of the workflow, corresponding to matrix assembly and preconditioner setup, is executed on the CPU, preventing further acceleration using GPUs. Major concerns~\cite{b37} with the existing GPU interface libraries and creating a direct GPU-port of OpenFOAM are:
\begin{itemize}
  \item Existing third-party libraries and packages have optimized GPU-code, however the interface can provide acceleration to only a limited part of the computation, as seen with PETSc interface in Figure~\ref{fig:openfoam-time-step-details}.
  \item Code duplication: OpenFOAM has $\sim O$(1M) lines of code, and porting and maintaining a separate/duplicate GPU-code will be tedious.
  \item Vendor-specific GPU code and proprietary packages can limit performance-portability, thereby restricting users to a specific hardware type.
  \item Code ownership and maintenance can be challenging, especially with vendor-specific code. Open-source projects thrive on community engagement, but lack of vendor-specific GPU programming knowledge can hinder involvement and slow down application development.
\end{itemize}

\begin{figure}[ht]
  \centerline{\includegraphics[width=0.5\textwidth]{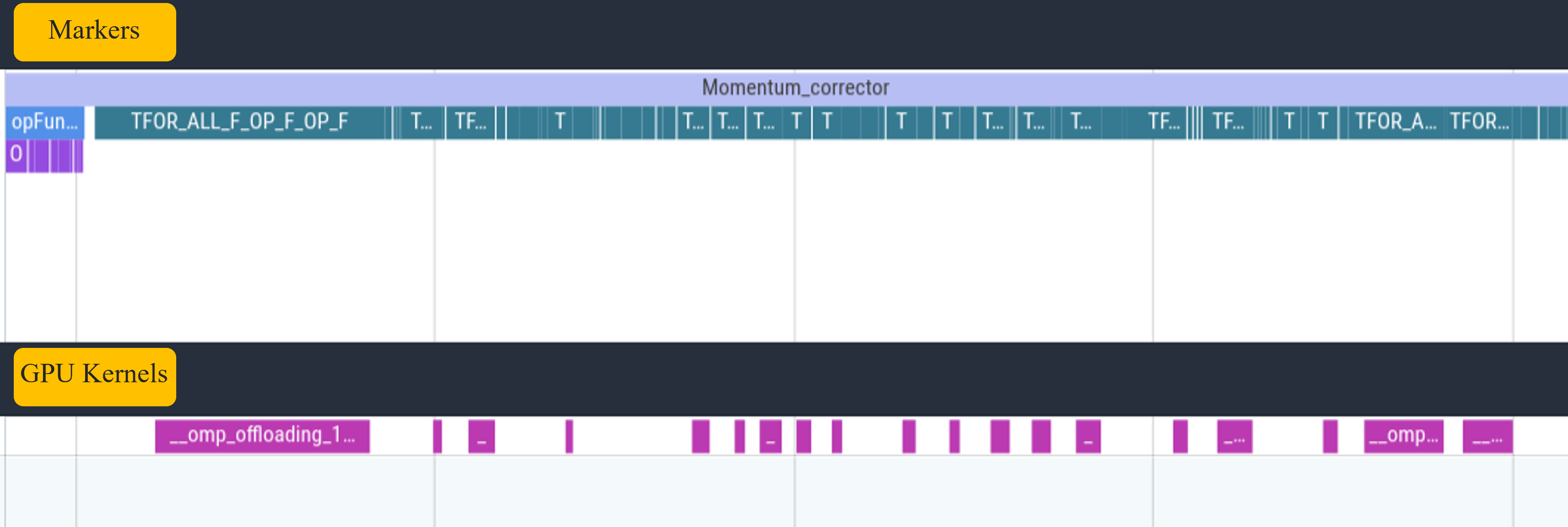}}
  \caption{A section of the trace collected of {\tt simpleFoam} solver showing offloading of macro expressions, here {\tt TFOR\_ALL\_F\_OP\_F\_OP\_F} in the momentum corrector stage, on MI300A.}
  \label{fig:openfoam-macro}
\end{figure}

To this effect, our approach for accelerating OpenFOAM is fundamentally different, where we: i) embrace the unified memory model, ii) identify {\it for loops} which can be parallelized (sometimes using atomics) and add a single line of compiler directives to express parallelism and allow multithreaded code execution on compute elements of the CPU or the GPU, and iii) accelerate code regions that closely follow each other and share the data. The last point deserves additional explanation. Considering that our target architecture supports unified physical memory, and considering that frequently alternating the execution between the CPU and the GPU carries very small performance penalty, we initially take a rather opportunistic approach where we offload the ``low hanging fruits" with maximum impact on the performance. For example, line 27 of listing~\ref{lst:openfoam-simple} uses operator overloading and macro expressions as shown in listing~\ref{lst:openfoam-tforall} to perform field operations for momentum correction. The {\tt for} loops in macro expressions are offloaded with OpenMP compiler directives and are called multiple times, as seen in the trace in Figure~\ref{fig:openfoam-macro}, thereby resulting in substantial acceleration. Simply put, adding a single line with OpenMP directives results in offloading significantly large number of loops within each time step. Note the use of {\tt if(target:n>TARGET\_CUT\_OFF)} construct allows adaptive switching of the execution between the CPU cores and the GPU. This construct is very useful on an APU where switching between the computation on the host and device has a low overhead.

\begin{lstlisting}[caption={Code snippet from {\tt FieldM.H} showing directive-based offloading of macro expressions.}, language=C++, label={lst:openfoam-tforall}]
#define TPARALLELFOR_ALL_F_OP_F_OP_F(typeF1, f1, OP1, typeF2, f2, OP2, typeF3, f3)     \
                                              \
    /* Check fields have same size */         \
    checkFields(f1, f2, f3, "f1 " #OP1 " f2 " #OP2 " f3");                                \
                                          \
    /* Field access */                    \
    List_ACCESS(typeF1, f1, f1P);         \
    List_CONST_ACCESS(typeF2, f2, f2P);   \
    List_CONST_ACCESS(typeF3, f3, f3P);   \
    label loop_len = (f1).size();         \
                                          \
    /* Loop: f1 OP1 f2 OP2 f3 */          \
     _Pragma("omp target teams distribute parallel for if(target:loop_len > TARGET_CUT_OFF)")  \
    for (label i = 0; i < loop_len; ++i) \
    {                                    \
        (f1P[i]) OP1 (f2P[i]) OP2 (f3P[i]);  \
    }
\end{lstlisting}

We further add OpenMP directives to other parallelizable loops. In {\tt simpleFoam} (listing~\ref{lst:openfoam-simple}), the {\tt solve()} routine on line 10 solves the linear system of equations to predict the velocity. Listing~\ref{lst:openfoam-pbicgstab} shows a snippet from the {\tt PBiCGStab} solver used to solve the momentum equations in the {\tt HPC\_motorbike} benchmark. The solver inturn invokes the preconditioner (line 4) and an example code is shown in listing~\ref{lst:openfoam-dilu}. Both the solver and preconditioner implementations have {\tt for} loops that iterate over the available cells (or discrete volumes) $\displaystyle \partial V_i$. OpenMP is used to offload these regions with {\tt target} clause and {\tt teams distribute parallel for} compiler optimizations. 

On systems supporting unified virtual memory addressing, the OpenMP pragma {\tt unified\_shared\_memory} is used to simplify data management across CPU and GPU. The OpenMP runtime pass host pointers by-value to the device (see Section \ref{sec:MI300A}), which provides the ease and flexibility to offload complex regions of code. On MI300A, which supports both the unified virtual addressing and unified physical memory addressing, {\tt unified\_shared\_memory} results in a highly performant implementation.   

\begin{lstlisting}[caption={Code snippet from {\tt PBiCGStab.C} showing directive-based offloading on MI300A.}, language=C++, label={lst:openfoam-pbicgstab}]
  #pragma omp requires unified_shared_memory
  ...
  // --- Precondition pA
  preconPtr->precondition(yA, pA, cmpt);
  ...
  // --- Calculate sA
  #pragma omp target teams distribute parallel for if(target:nCells>TARGET_CUT_OFF)
  for (label cell=0; cell<nCells; cell++)
  {
      sAPtr[cell] = rAPtr[cell] - alpha*AyAPtr[cell];
  }
  ...
\end{lstlisting}

\begin{lstlisting}[caption={Code snippet from {\tt DILUPreconditioner.C} showing directive-based offloading on MI300A.}, language=C++, label={lst:openfoam-dilu}]
  #pragma omp requires unified_shared_memory
  ...
  #pragma omp target teams distribute parallel for if(target:nCells>TARGET_CUT_OFF)
  for (label cell=0; cell<nCells; cell++)
  {
      wAPtr[cell] = rDPtr[cell]*rAPtr[cell];
  }
  ...
\end{lstlisting}

\begin{figure*}[t]
  \centerline{\includegraphics[width=1.0\textwidth]{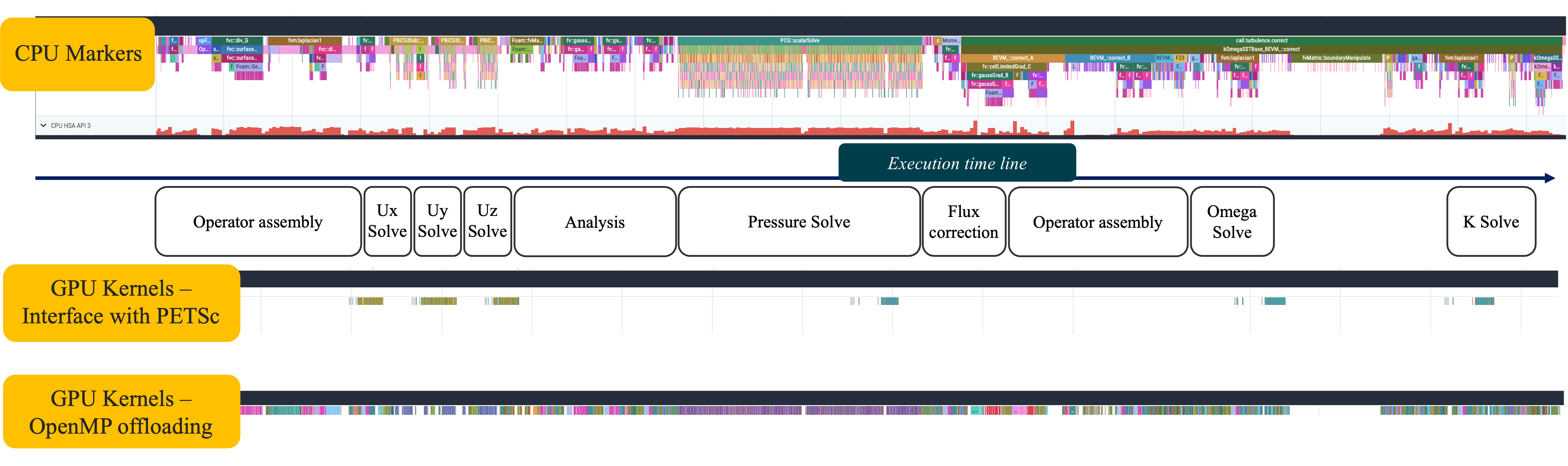}}
  \caption{A trace of the {\tt HPC\_motorbike} benchmark (Large with 34 M cells) using {\tt simpleFoam} solver showing directive-based offloading with OpenMP and {\tt unified\_shared\_memory} on MI300A. In comparison to the GPU-acceleration achieved with PETSc interface (middle), directive-based offloading with OpenMP 5.0 (bottom) is able to offload significantly larger portions of the code to MI300A, thereby resulting is significantly higher acceleration than that offered by PETSc and other third-party interfaces. For simplicity, the trace included here contains details of a single time step only.}
  \label{fig:openfoam-openmp}
\end{figure*}

To this end, the key advantages of using directive-based approach in OpenFOAM and unified physical memory
are: 
\begin{itemize}
  \item  Figure~\ref{fig:openfoam-openmp} shows a trace of {\tt HPC\_motorbike} benchmark on MI300A with OpenMP target offloading and \\ {\tt unified\_shared\_memory}. With $O$(100) lines of code modification, substantially more code than in OpenFoam-to-PETSc (Figure~\ref{fig:openfoam-time-step-details}) approach is offloaded to the device, resulting in better device utilization and increased speedups. Accordingly, the time and effort taken to port and tune production-ready code is significantly lower than conventional GPU-programming with HIP or CUDA and discrete memory spaces.
  \item Unified memory reduces the complexities around data management. All of {\tt simpleFoam} (listing~\ref{lst:openfoam-simple}) was offloaded without the need to add explicit {\tt map} clause and manage data environment (see OpenMP data environments in\cite{b38}). To reduce the number of code lines and to simplify the porting effort, we heavily relied on the {\it implicit declare target} feature that we explained in Section \ref{sec:MI300A} as many variables in the body of the {\it for loops} are complex objects (classes) using overloaded operators.
  \item Unified memory keeps the total memory footprint almost identical to the original CPU-only code, which is in stark contrast to using third-party interfaces and libraries like PETSc, where additional CPU memory is needed for conversion between matrix formats and GPU memory is needed for data replication. For example, the {\tt HPC\_motorbike} benchmark requires more than 80 GB of GPU memory and consumes almost 2x more system memory, compared to the non-accelerated OpenFOAM.
  \item Minimal code duplication, and high portability. On HPC systems where either the accelerators/GPUs are not available or the compiler lacks support for target offloading, multi-thread parallelism on CPU cores is obtained with the same compiler directives.
  \item The {\tt if(target:n$>$TARGET\_CUT\_OFF)} feature in OpenMP allows  the runtime to dispatch the computations to the desired hardware based on the iteration count {\tt TARGET\_CUT\_OFF} as shown in listings~\ref{lst:openfoam-tforall}, \ref{lst:openfoam-pbicgstab}, and \ref{lst:openfoam-dilu}.
\end{itemize}

To our knowledge, our work is the first demonstration of the potential of the APU,  OpenMP standard, and unified memory to accelerate a complex production application with relatively low effort. There is an ongoing effort to upstream our unified memory implementation in OpenFOAM (currently accessible through GitHub~\cite{b39}) to official repository~\cite{b40}.

% ==============================================
% SECTION V: Performance Evaluation: The APU Advantage
% ==============================================
\section{Performance Evaluation: The APU Advantage}
\label{sec:performance}

The performance of directive-based offloading in OpenFOAM with {\tt unified\_shared\_memory} is assessed by running the {\tt HPC\_motorbike} benchmark (Large with 34 M cells) on a single MI300A. The benchmark is configured to run for 20 time-steps, with the average time of execution per time-step (in sec.) taken as the figure of merit (FOM). Performance of the APU is compared to that of the systems with discrete CPU (dCPU) and discrete GPU (dGPU: MI210, A100, and H100). The high level system details are provided in Table~\ref{tab:platforms}. All systems must be configured and enabled with heterogeneous memory management (HMM) to allow the GPU to address the system memory and accordingly run OpenFOAM with OpenMP offloading and {\tt unified\_shared\_memory}. Furthermore, memory pooling is employed to improve performance by reusing the allocated memory (for buffers larger than 5K elements) instead of frequently allocating and deallocating memory. An interface with the Umpire library~\cite{b41} allocates and provides the memory pool using different memory allocators --- HIP/CUDA Managed Memory provides an optimal solution for systems with dGPU platforms~\cite{b42}, whereas any memory allocator can be used with MI300A.

\begin{figure}[t]
  \centerline{\includegraphics[width=0.35\textwidth]{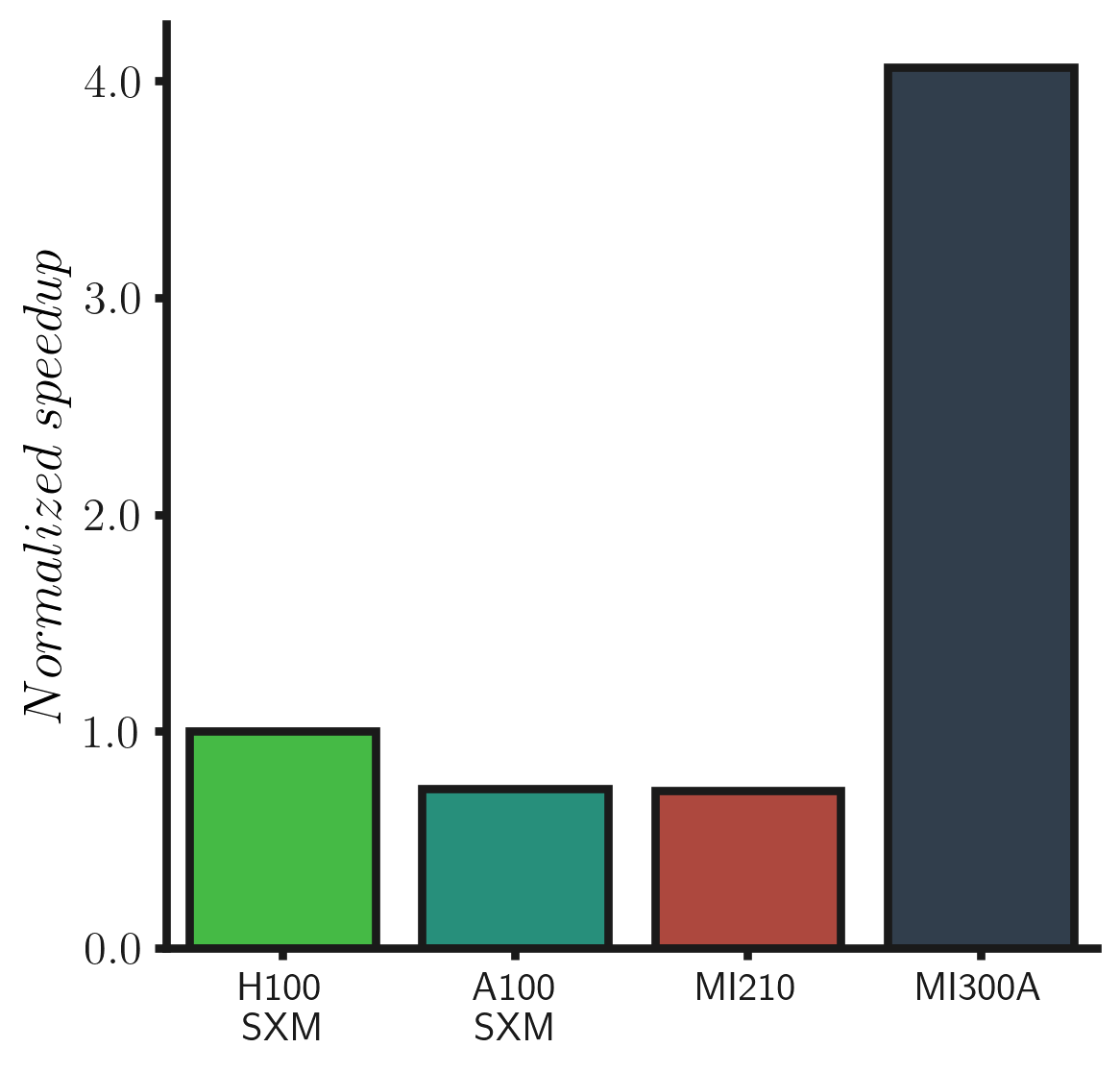}}
  \caption{Normalized speedup for the {\tt HPC\_motorbike} benchmark (Large with 34 M cells) on different devices with respect to the observed FOM on an x86-based H100-SXM. The APU offers significant performance gain over discrete GPUs.}
  \label{fig:openfoam-speedups}
\end{figure}

\begin{figure}[t]
  \centerline{\includegraphics[width=0.45\textwidth]{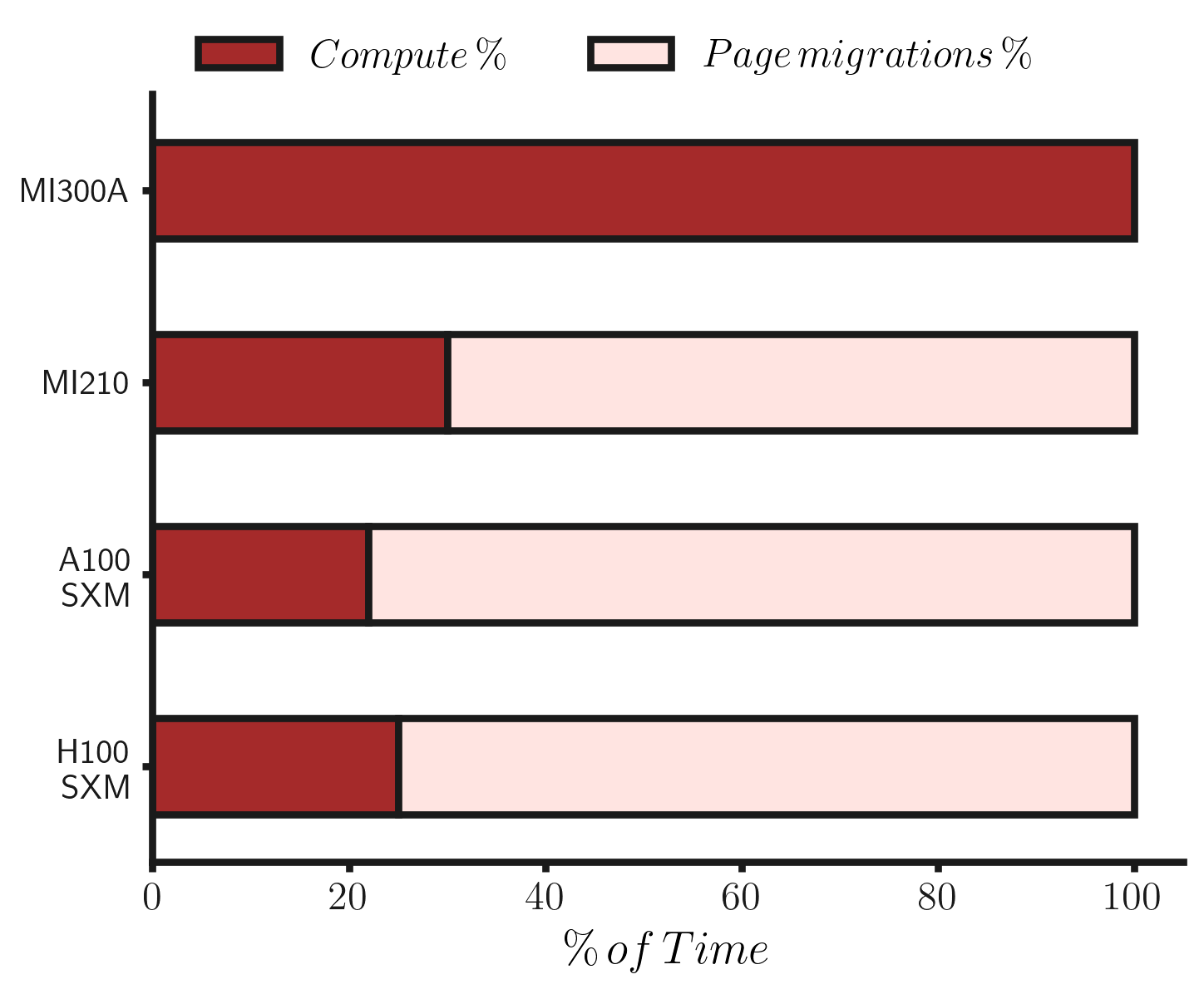}}
  \caption{Profiling analysis from different platforms showing the portion of time spent in page migrations (updating GPU tables and copying the data between host and device) during the execution of the {\tt HPC\_motorbike} benchmark (Large with 34 M cells). The unified physical memory shared between the CPU and compute cores of the APU removes the overhead of page migrations, providing significant acceleration in comparison to discrete GPUs.}
  \label{fig:openfoam-profiles}
\end{figure}

Normalized speedups with respect to the FOM observed on a system with x86 dCPU and H100-SXM dGPU are shown in Figure~\ref{fig:openfoam-speedups}. On each system, a single CPU-core (and single process) is used to offload work to the device and the benchmark is run five times to record the average FOM. MI300A offers significant performance gain over systems with dGPUs with 4x speedup over a system with H100-SXM and 5x speedup over the previous generation MI210. Profiling analysis (Figure~\ref{fig:openfoam-profiles}) reveals that:
\begin{itemize}
	\item On dGPUs, more than $65\%$ of the time is spent in page migrations: updating GPU tables and copying the data between host and device.
	\item On APU, the unified physical memory shared between the CPU cores and GPU's compute units (Figure~\ref{fig:dgpu-vs-apu}) completely removes the overhead of page migrations, resulting in significant performance boost.
\end{itemize}

\begin{table*}[t]
  \caption{Details of different discrete GPU platforms used to compare with MI300A.}
  \centering \ra{1.3} 
  \begin{tabular}{lllcc}
  \toprule
  \rowcolor{cyan}
  Device 	     & Stack, Compiler, NPS          				 				& Memory Pool    \\ \midrule
  MI300A	     				& ROCm-6.0, {\tt amdclang++ (clang-17.0)}, NPS1 & System memory  \\
  x86, MI210, PCIe 4.0		    & ROCm-6.0, {\tt amdclang++ (clang-17.0)}, NPS1 & Managed memory  \\
  x86, A100-80GB SXM, PCIe 4.0  & CUDA-12.2.2, {\tt clang-18.0}, NPS1           & Managed memory \\
  x86, H100-SXM, PCIe 5.0       & CUDA-12.2.2, {\tt clang-18.0}, NPS1           & Managed memory \\ \bottomrule           
  \end{tabular}
  \label{tab:platforms}
\end{table*}

Measurements were collected on single-socket (64-cores) AMD EPYC$^\text{TM}$ ``Zen"4 CPU by running the same {\tt HPC\_motorbike} benchmark and OpenFOAM source. Numerical experiments confirm that 1 MI300A with 1 CPU-core has 2x better performance than a single-socket ``Zen"4 CPU. Furthermore, overloading MI300A with multiple processes improves performance, for example, tests with 3-6 CPU-cores per APU gives 2x better performance than reported in Table~\ref{tab:platforms}. Whereas, with unified memory, overloading a device in dCPU+dGPU system gives marginal gain in performance due to lack of scalability and serialization of certain parts of page migration and page registration within the OS image. For simplicity, Figures~\ref{fig:openfoam-speedups} and \ref{fig:openfoam-profiles} compare the performance of GPUs only with emphasis on speedup of one APU over a single dGPU, and thus the CPU-only execution and overloading an APU with multiple CPU-cores is omitted from further discussion.

% ==============================================
% SECTION VI: Summary/conclusions
% ==============================================
\section{Conclusions}
\label{sec:conclusion}

In this paper we discuss OpenMP target offloading and implementation of {\tt unified\_shared\_memory} in the AMD ROCm$^\text{TM}$ OpenMP compiler. A blueprint of the APU programming model is presented, which leverages unified memory on MI300A, and key distinctions compared to the OpenMP target offloading on discrete GPUs are highlighted. OpenFOAM, an open-source C++ library for computational fluid dynamics, is ported to MI300A with minimal code modifications and the addition of OpenMP and {\tt unified\_shared\_memory} demonstrates the ease and flexibility of porting production-ready codes to MI300A using OpenMP directives.

\section*{Acknowledgment}

The authors gratefully acknowledge the OpenFOAM HPC Technical Committee, in particular, Ivan Spisso and Stefano Zampini, and Michael Klemm from AMD's HPC Center of Excellence for the frequent discussions and insights on OpenFOAM architecture and code design strategies. We also extend our gratitude to Rajneesh Bhardwaj and Felix Kuehling from the AMD Kernel team for their invaluable support.

AMD, the AMD Arrow logo, Instinct, EPYC, CDNA, ROCm, and combinations thereof are trademarks of Advanced Micro Devices, Inc. Other product names used in this publication are for identification purposes only and may be trademarks of their respective companies.


\begin{thebibliography}{00}
\bibitem{b1} Cardwell, S.G., Vineyard, C., Severa, W., Chance, F.S., Rothganger, F., Wang, F., et al., Truly heterogeneous HPC: Co-design to achieve what science needs from HPC. In {\it 17th Smoky Mountains Comp. Sciences and Eng. Conf., SMC 2020}, August, 2020, pp. 349--365.
\bibitem{b2} Chai, L., Gao, Q. and Panda, D. K., Understanding the impact of multi-core architecture in cluster computing: A case study with intel dual-core system. In {\it 17th IEEE Int. Symp. Cluster Comp. and the Grid (CCGrid'07)}, May, 2007, pp. 471--478.
\bibitem{b3} The Top500 List (https://www.top500.org/lists/top500/), {\it Accessed}, July 2023.
\bibitem{b4} Gao, Y. and Zhang, P., A survey of homogeneous and heterogeneous system architectures in high performance computing. In {\it 2016 IEEE Int. Conf. Smart Cloud (SmartCloud)}, November, 2016, pp. 170--175.
\bibitem{b5} Bak, S., Bertoni, C., Boehm, S., Budiardja, R., Chapman, B.M., Doerfert, J., et al., OpenMP application experiences: Porting to accelerated nodes. {\it Parallel Computing}, 2022, vol 109, pp. 102856.
\bibitem{b6} OpenMP\textregistered\ (https://www.openmp.org)
\bibitem{b7} OpenACC (https://www.openacc.org)
\bibitem{b8} OpenMP\textregistered\ 4.0 Specifications, 2013 (https://www.openmp.org/wp-content/uploads/OpenMP4.0.0.pdf)
\bibitem{b9} OpenMP\textregistered\ 5.0 Specifications, 2018 (https://www.openmp.org/wp-content/uploads/OpenMP-API-Specification-5.0.pdf)
\bibitem{b10} Nompelis, I., Jost, G., Koniges, A., Daley, C., Eder, D. and Stone, C., Enabling Execution of a Legacy CFD Mini Application on Accelerators Using OpenMP. In {\it High Perf. Comp., ISC High Performance 2020}, June, 2020, Proceedings 35, pp. 270--287.
\bibitem{b11} Daley, C., Ahmed, H., Williams, S. and Wright, N., A case study of porting HPGMG from CUDA to OpenMP target offload. In {\it 16th Int. Workshop on OpenMP, IWOMP 2020}, September, 2020, Proceedings 16, pp. 37--51.
\bibitem{b12} Mishra, A., Li, L., Kong, M., Finkel, H. and Chapman, B., Benchmarking and evaluating unified memory for OpenMP GPU offloading. In {\it 4th Workshop on the LLVM Compiler Infrastructure in HPC}, November 2017, pp. 1--10.
\bibitem{b13} Martineau, M. and McIntosh-Smith, S., The productivity, portability and performance of OpenMP 4.5 for scientific applications targeting Intel CPUs, IBM CPUs, and NVIDIA GPUs. In {\it 13th Int. Workshop on OpenMP, IWOMP 2017}, September 20–22, 2017, Proceedings 13, pp. 185--200.
\bibitem{b14} Weller, H.G., Tabor, G., Jasak, H. and Fureby, C., A tensorial approach to computational continuum mechanics using object-oriented techniques. {\it Computers in Physics}, 1998, vol 12(6), pp.620--631.
\bibitem{b15} HIP Runtime (https://rocm.docs.amd.com/en/latest/reference/hip.html).
\bibitem{b16} CUDA\textregistered\ Toolkit (https://developer.nvidia.com/cuda-toolkit).
\bibitem{b17} Kim, J., Baczewski, A.D., Beaudet, T.D., Benali, A., Bennett, M.C., Berrill, M.A., et al., QMCPACK: an open source ab initio quantum Monte Carlo package for the electronic structure of atoms, molecules and solids. {\it J. Phy.: Condensed Matter}, 2018, vol 30(19), pp. 195901. 
\bibitem{b18} Maintz, S. and Wetzstein, M., Strategies to accelerate VASP with gpus using OpenACC. In {\it Proceedings of the Cray User Group}, 2018.
\bibitem{b19} Sawyer, W., Zaengl, G. and Linardakis, L., Towards a multi-node OpenACC Implementation of the ICON Model. In {\it EGU Gen. Assembly Conf.}, 2014, pp. 15276.
\bibitem{b20} Budiardja, R.D. and Cardall, C.Y., Targeting GPUs with OpenMP directives on Summit: A simple and effective Fortran experience. {\it Parallel Comp.}, 2019, vol 88, pp.102544.
\bibitem{b21} Hart, A., First experiences porting a parallel application to a hybrid supercomputer with OpenMP4. 0 device constructs. In {\it Int. Workshop on OpenMP}, October 2015, pp. 73--85. 
\bibitem{b22} Karlin, I., Scogland, T., Jacob, A.C., Antao, S.F., Bercea, G.T., Bertolli, C., et al., Early experiences porting three applications to OpenMP 4.5. In {\it 12th Int. Workshop on OpenMP, IWOMP 2016}, October 5-7, 2016, Proceedings 12, pp. 281--292.
\bibitem{b23} Martineau, M., McIntosh-Smith, S. and Gaudin, W., Evaluating OpenMP 4.0's effectiveness as a heterogeneous parallel programming model. In {\it 2016 IEEE Int. Parallel and Distributed Processing Symposium Workshops (IPDPSW)}, May 2016, pp. 338--347.
\bibitem{b24} Hayashi, A., Shirako, J., Tiotto, E., Ho, R. and Sarkar, V., Performance evaluation of OpenMP's target construct on GPUs-exploring compiler optimisations. {\it Int. J. High Performance Computing and Networking}, 2019, vol 13(1), pp.54--69.
\bibitem{b25} Martineau, M., Price, J., McIntosh-Smith, S. and Gaudin, W., Pragmatic performance portability with OpenMP 4. x. In {\it 12th Int. Workshop on OpenMP, IWOMP 2016}, October 5-7, 2016, Proceedings 12, pp. 253--267.
\bibitem{b26} Tiotto, E., Mahjour, B., Tsang, W., Xue, X., Islam, T. and Chen, W., OpenMP 4.5 compiler optimization for GPU offloading. {\it IBM J. Research and Development}, 2019, vol 64(3/4), pp.14--1.
\bibitem{b27} Doerfert, J., Diaz, J.M.M. and Finkel, H.The TRegion interface and compiler optimizations for OpenMP target regions. In {\it 15th Int. Workshop on OpenMP, IWOMP 2019}, September 11–13, 2019, Proceedings 15, pp. 153--167.
\bibitem{b28} Diaz, J.M., Friedline, K., Pophale, S., Hernandez, O., Bernholdt, D.E. and Chandrasekaran, S. Analysis of OpenMP 4.5 offloading in implementations: correctness and overhead.{\it Parallel Computing}, 2019, vol 89, pp .102546.
\bibitem{b29} Grinberg, L., Bertolli, C. and Haque, R., Hands on with OpenMP4. 5 and unified memory: developing applications for IBM’s hybrid CPU+ GPU systems (Part I). In {\it 13th Int. Workshop on OpenMP, IWOMP 2017}, September 20–22, 2017, Proceedings 13, pp. 3--16.
\bibitem{b30} Edwards, H.C., Trott, C.R. and Sunderland, D., Kokkos: Enabling manycore performance portability through polymorphic memory access patterns. {\it J. Parallel and Distributed Computing}, 2014, vol 74(12), pp.3202--3216.
\bibitem{b31} Grinberg, L., Bertolli, C. and Haque, R., Hands on with OpenMP4. 5 and unified memory: developing applications for IBM’s hybrid CPU+ GPU systems (Part II). In {\it 13th Int. Workshop on OpenMP, IWOMP 2017}, September 20–22, 2017, Proceedings 13, pp. 17--29.
\bibitem{b32} Jasiński, D., Adapting OpenFOAM for massively parallel GPU architecture. In {\it 3rd OpenFOAM User Conference}, 2015.
\bibitem{b33} Mills, R.T., Adams, M.F., Balay, S., Brown, J., Dener, A., Knepley, M., et al., Toward performance-portable PETSc for GPU-based exascale systems. {\it Parallel Computing}, 2021, vol 108, pp.102831.
\bibitem{b34} Cojean, T., Tsai, Y.H.M. and Anzt, H., Ginkgo—A math library designed for platform portability. {\it Parallel Computing}, 2022, vol 111, pp .102902.
\bibitem{b35} Caretto, L.S., Gosman, A.D., Patankar, S.V. and Spalding, D.B., Two calculation procedures for steady, three-dimensional flows with recirculation. In {\it 3rd Int. Conf. Numerical Methods in Fluid Mechanics: Vol. II Problems of Fluid Mechanics}, 1973, pp. 60--68.
\bibitem{b36} OpenFOAM\textregistered\ HPC Benchmark Suite \\(https://develop.openfoam.com/committees/hpc/).
\bibitem{b37} OpenFOAM\textregistered\ HPC Technical Committee (https://wiki.openfoam.com).
\bibitem{b38} Technical Report 11: First preview for the OpenMP\textregistered API Version 6.0 – November 2022 (https://www.openmp.org/wp-content/uploads/openmp-TR11.pdf)
\bibitem{b39} OpenFOAM with OpenMP (https://github.com/ROCm/OpenFOAM\_HMM).
\bibitem{b40} OpenFOAM Source (https://develop.openfoam.com/Development/openfoam).
\bibitem{b41} Beckingsale, D., Mcfadden, M., Dahm, J., Pankajakshan, R. and Hornung, R., Umpire: Application-Focused Management and Coordination of Complex Hierarchical Memory, in {\it IBM J. Research and Development}.
\bibitem{b42} Hubbard, J., Brito, G., Garg, C., Sakharnykh, N. and Oh, F., Simplifying GPU Application Development with Heterogeneous Memory Management, 2023 (https://developer.nvidia.com/blog/simplifying-gpu-application-development-with-heterogeneous-memory-management/)
\end{thebibliography}
\end{document}